\documentclass[prb,preprint,superscriptaddress,groupedaddress,showpacs,preprintnumbers,amsmath,amssymb]{revtex4}

\usepackage{graphicx}
\usepackage{dcolumn}
\usepackage{bm}

\begin{document}


\title{Generalized band anti-crossing model for highly mismatched semiconductors applied to BeSe$_{x}$Te$_{1 - x}$}


\author{Titus Sandu}
 \email{sandu@asu.edu}
\affiliation{%
Chemical and Materials Engineering Department, \\
 Arizona State University, Tempe, AZ 85287-6006\\}%
\author{W. P. Kirk}%
\affiliation{%
NanoFAB Center, Electrical Engineering Department,\\
   University of Texas at Arlington, Arlington, TX 76019\\}%



\date{\today}

\begin{abstract}
We report a new model for highly mismatched semiconductor 
(HMS) alloys. Based on the Anderson impurity Hamiltonian, the model 
generalizes the recent band anti-crossing (BAC) model, which successfully 
explains the band bowing in highly mismatched semiconductors. Our model is 
formulated in empirical tight-binding (ETB) theory and uses the so called 
sp$^{3}$s* parameterization. It does not need extra parameters other than 
bulk ones. The model has been applied to BeSe$_{x}$Te$_{1 - x}$ alloy. BeTe 
and BeSe are wide-band gap and highly mismatched semiconductors. 
Calculations show large band bowing, larger on the Se rich side than on the 
Te rich side. Linear interpolation is used for an arbitrary concentration 
$x$. The results are applied to calculation of electronic and optical 
properties of BeSe$_{0.41}$Te$_{0.59}$ lattice matched to Si in a 
superlattice configuration. 

\end{abstract}


\maketitle


\section{Introduction}

In recent years there has been a tremendous interest in electro-optical 
properties of II-VI semiconductors in particular epitaxial II-VI 
heterostructures. New developments \cite{Clark:2000,Kirk:2000} in 
the growth of Si lattice-matched BeSe$_{0.41}$Te$_{0.59}$ open the 
opportunity to a new class of Si based devices. Be-chalcogenides are 
wide-band gap zinc blende semiconductors with lattice constants close to 
that of Si. Thus BeTe and BeSe have the lattice constants of 5.6269 and 
5.1477 {\AA}, respectively, 3.6 {\%} larger and 5.2 {\%} smaller than Si. 
Vegard's law says that the lattice matched composition with Si is 
BeSe$_{0.41}$Te$_{0.59}$. Therefore Be-chalcogenides are candidates for 
Si-based heterostructures. 

The difference in size and orbital energies 
between Se and Te in addition to large lattice mismatch between BeTe and 
BeSe makes the virtual-crystal approximation (VCA) inappropriate for the 
ternary alloy BeSe$_{x}$Te$_{1 - x}$. The band anti-crossing (BAC) model 
\cite{Walukiewicz:2000} has been introduced in order to explain the 
electronic structure of highly mismatched III-V-N alloys and II-VI alloys 
like ZnSe$_{x}$Te$_{1 - x}$ \cite{Walukiewicz:2003}. At impurity like 
concentrations close to both end points, the electronegativity difference 
between constituent elements gives rise to localized energy levels close to 
the conduction or valence band. Thus in the ZnSe(ZnTe)-rich side, the band 
gap bowing is mostly determined by the anticrossing interaction between the 
Te(Se) localized level, which behaves like an impurity, and the extended 
states of ZnSe valence band (ZnTe conduction band) near the center of the 
Brillouin zone \cite{Walukiewicz:2004}. 

In this communication we develop 
a model which is a natural extension of the BAC model to empirical 
tight-binding (ETB). Based on this model we determine optical bandgap of 
BeSe$_{x}$Te$_{1 - x}$ alloy and further we analyze the band folding in 
Si/BeSe$_{0.41}$Te$_{0.59}$ heterostructures.

\section{Model}

There are several studies employing tight-binding (TB) models for HMS. They 
use either supercells \cite{Reilly:2002} or add extra parameters to the 
usual TB parameters within the BAC model \cite{Shtinkov:2003}. Our 
approach needs no extra parameters others than the usual TB parameters and 
is a natural extension of the BAC model to TB. We consider first the dilute 
limit. The starting point is the impurity model of Anderson 
\cite{Anderson:1961}. Consider a complex-structured impurity interacting 
with the host crystal having the Hamiltonian

\begin{equation}
\label{eq1}
H = H_c + H_{imp} + V = H_0 + V
\end{equation}

\noindent
where $H_{c}$ is Hamiltonian of the host crystal, $H_{imp}$ is the Hamiltonian 
of the impurity and $V$ is the Hamiltonian of the interaction between impurity 
and crystal. The above Hamiltonians have the following expressions
\begin{equation}
\label{eq1a}
H_{imp} = \sum\limits_{p,d,\sigma } {\varepsilon _{d\sigma } c_{pd\sigma }^ 
+ c_{pd\sigma } } ,
\end{equation} 
\begin{equation}
\label{eq1b}
H_c = \sum\limits_{i,k,\sigma } {\varepsilon 
_{ik\sigma } c_{ik\sigma }^ + c_{ik\sigma } }  ,
\end{equation}
\begin{equation}
\label{eq1c}
V = \sum\limits_{p,d,i,k,\sigma } {\left( {V_{ikd} c_{pd\sigma }^ + 
c_{ik\sigma } + h.c.} \right)} . 
\end{equation}
Here \textit{$\sigma $} is the spin index, $d$ are the energy levels of impurity at the $p$th site 
($N_{i }$, the total number of impurities), $i$ is band index of the crystal, $k$ is 
the wavevector, $c_{pd\sigma }^ + (c_{pd\sigma } )$are the creation 
(annihilation) operators of electrons on impurity levels. $c_{ik\sigma }^ + 
(c_{ik\sigma } )$ are the creation (annihilation) operators of electrons of 
the bands in the crystal, and $V$ is the coupling between impurity and 
crystal and has factor $1 \mathord{\left/ {\vphantom {1 {\sqrt N }}} \right. 
\kern-\nulldelimiterspace} {\sqrt N }$ with $N$ total number of sites. In order 
to calculate the effect of impurity we shall use an expression for the 
projection of $H$ into \textbf{${\rm H}$}, a subspace of the Hilbert space 
spanned by the Hamiltonian $H_{0}$. We define $P$ (and $Q)$ the projections onto 
(out of) \textbf{${\rm H}$ }as $P^2 = P$, $Q^2 = Q$, $P + Q = 1$, $PQ = 
0$, $\left[ {P,H_0 } \right] = \left[ {Q,H_0 } \right] = 0$. Let the Green 
function of the entire system be denoted by $G\left( z \right) = \left( {z - 
H} \right)^{ - 1}$ and the projected part onto \textbf{${\rm H}$} as 
$\overline G \left( z \right) = PG\left( z \right)P$. In this way we denote 
the projection onto \textbf{${\rm H}$} of any operator $A$ as: $\overline A 
= PAP$. Straightforward algebra gives us

\begin{equation}
\label{eq2}
\overline G \left( z \right) = \frac{1}{z - \overline H _0 - \overline R 
\left( z \right)},
\end{equation}

\noindent
where $\overline R \left( z \right) = PVP + PV\frac{Q}{z - H_0 - QVQ}VP$. One 
can expand $\overline R \left( z \right)$ for small $V$

\begin{equation}
\label{eq3}
\overline R \left( z \right) = PVP + PV\frac{Q}{z - H_0 }VP + PV\frac{Q}{z - 
H_0 }V\frac{Q}{z - H_0 }VP + \ldots  .
\end{equation}

\noindent
Due to $Q$ in the expansion, all the intermediate states are outside of 
\textbf{${\rm H}$}, therefore the diagrams representing $\overline R \left( 
z \right)$ must be irreducible. If $\overline R \left( z \right)$ is small 
compared to $H_c $, Eq. (\ref{eq2}) can be expanded in a power series of $\overline 
R \left( z \right)$ as

\begin{equation}
\label{eq4}
\overline G \left( z \right) = G_c + G_c \overline R G_c + G_c \overline R 
G_c \overline R G_c + \ldots  .
\end{equation}

No approximations have been made so far. If $V$ is small we can replace 
$\overline R \left( z \right)$ in Eq. (\ref{eq3}) by the first two terms and summing 
up all contributions (the first one is 0 in our model). By making such an 
approximation we, in fact, sum an entire class of diagrams. The impurity 
averaging is made by noticing that all macroscopically observables are 
self-averaging, i.e. they have asymptotically exact values in the 
thermodynamic limit \cite{Lifshits:1964}. In other words the average of 
the product of such quantities is equal (within asymptotic accuracy) to the 
product of their averages. Therefore the impurity averaging in Eq. (\ref{eq2}) is 
simply taken as averaging $\overline {R\left( z \right)} $ in Eq. (\ref{eq3}). By 
keeping the first two terms in Eq. (\ref{eq3}), the model is equivalent to the 
optical model laid out in Ref. \onlinecite{Yonezawa:1964}: it only brings the 
shift to the energy levels of the unperturbed Hamiltonian and gives the 
following form for the diagonal part of the Green function of $i$th band

\begin{equation}
\label{eq5}
G_{kk\sigma \sigma }^{ii} \left( z \right) = \frac{1}{1 - \varepsilon 
_{ik\sigma } - x\sum\limits_d {\frac{\left| {V_{ikd} } \right|^2}{z - 
\varepsilon _{d\sigma } }} },
\end{equation}

\noindent
where $x$ is the dilute concentration. For one band and 1-level impurity the 
result is identical to the BAC model \cite{Walukiewicz:2000,Walukiewicz:2003,Walukiewicz:2004,Shan:2002}. 
This result can be 
easily expanded to include pair impurity interactions in addition to single 
impurity interactions \cite{Fahy:2004}.

\section{Application to sp$^{3}$s$^{*}$ TB Hamiltonian and Numerical Results}
The model laid out in the preceding section is directly applied to 
sp$^{3}$s* Hamiltonian \cite{Vogl:1983} with spin-orbit interaction 
\cite{Chadi:1977} The TB Hamiltonian is written in the sp$^{3}$ hybrid 
basis and the basis is rotated in such a way that a unit cell is formed by 
the anion hybrid orbitals and the cation hybrid orbitals pointed toward the 
anion site. The s* orbitals remain unchanged. In the no spin-orbit case, the 
transformation is $H_{hyb} \left( k \right) = S^ + \left( k \right)H_{AO} 
\left( k \right)S\left( k \right)$, where $H_{AO} \left( k \right)$ is the 
TB Hamiltonian in atomic (L\"{o}dwin) orbital basis and $H_{hyb} 
\left( k \right)$ the Hamiltonian in sp$^{3}$ hybrid basis. The S matrix is 
block diagonal in anion/cation index and has the form for anion and cation 
site, respectively, as

\begin{equation}
\label{eq6}
\frac{1}{2}\left[ {{\begin{array}{*{20}c}
 1 \hfill & 1 \hfill & 1 \hfill & 1 \hfill & 0 \hfill \\
 1 \hfill & 1 \hfill & { - 1} \hfill & { - 1} \hfill & 0 \hfill \\
 1 \hfill & { - 1} \hfill & 1 \hfill & { - 1} \hfill & 0 \hfill \\
 1 \hfill & { - 1} \hfill & { - 1} \hfill & 1 \hfill & 0 \hfill \\
 0 \hfill & 0 \hfill & 0 \hfill & 0 \hfill & 1 \hfill \\
\end{array} }} \right],
\end{equation}

\begin{equation}
\label{eq8}
\frac{1}{2}\left[ {{\begin{array}{*{20}c}
 {e^{ - ik \cdot {\kern 1pt} d_1 }} \hfill & {e^{ - ik \cdot {\kern 1pt} d_2 
}} \hfill & {e^{ - ik \cdot {\kern 1pt} d_3 }} \hfill & {e^{ - ik \cdot 
{\kern 1pt} d_4 }} \hfill & 0 \hfill \\
 { - e^{ - ik \cdot {\kern 1pt} d_1 }} \hfill & { - e^{ - ik \cdot {\kern 
1pt} d_2 }} \hfill & {e^{ - ik \cdot {\kern 1pt} d_3 }} \hfill & {e^{ - ik 
\cdot {\kern 1pt} d_4 }} \hfill & 0 \hfill \\
 { - e^{ - ik \cdot {\kern 1pt} d_1 }} \hfill & {e^{ - ik \cdot {\kern 1pt} 
d_2 }} \hfill & { - e^{ - ik \cdot {\kern 1pt} d_3 }} \hfill & {e^{ - ik 
\cdot {\kern 1pt} d_4 }} \hfill & 0 \hfill \\
 { - e^{ - ik \cdot {\kern 1pt} d_1 }} \hfill & {e^{ - ik \cdot {\kern 1pt} 
d_2 }} \hfill & {e^{ - ik \cdot {\kern 1pt} d_3 }} \hfill & { - e^{ - ik 
\cdot {\kern 1pt} d_4 }} \hfill & 0 \hfill \\
 0 \hfill & 0 \hfill & 0 \hfill & 0 \hfill & 1 \hfill \\
\end{array} }} \right].
\end{equation}

\noindent
The vectors in (\ref{eq8}) are ($a$ is the lattice constant): $d_1 = \left( {a 
\mathord{\left/ {\vphantom {a 4}} \right. \kern-\nulldelimiterspace} 4} 
\right)\left( {1,1,1} \right)$,$d_2 = \left( {a \mathord{\left/ {\vphantom 
{a 4}} \right. \kern-\nulldelimiterspace} 4} \right)\left( {1,\overline 1 
,\overline 1 } \right)$,$d_3 = \left( {a \mathord{\left/ {\vphantom {a 4}} 
\right. \kern-\nulldelimiterspace} 4} \right)\left( {\overline 1 
,1,\overline 1 } \right)$, and $d_4 = \left( {a \mathord{\left/ {\vphantom 
{a 4}} \right. \kern-\nulldelimiterspace} 4} \right)\left( {\overline 1 
,\overline 1 ,1} \right)$. An impurity is such a cell interacting through 
hybrid orbitals with the average/host crystal. 

BeSe and BeTe are quite new materials. The nearest-neighbor sp$^{3}$s* 
parameters were fitted to GW calculations \cite{Fleszar:2000}. They 
reproduce valence band edges $\Gamma _8 $ and $\Gamma _7 $, and the 
conduction band edges $\Gamma _6 $ and $X_{1}$. The average crystal is 
considered by the average parameters. The hopping parameters were scaled 
according to the Harrison scaling rule \cite{Harrison:1989} and then 
averaged. The band offset between BeTe and BeSe is considered to be 0.41 eV 
as indicated in Ref.\onlinecite{Bernardini:2000}. In this way one calculates the 
bowing of the on-site energies in addition to linear terms given by VCA. The 
hybrid states of the impurity lay outside bandgap of the host crystal, such 
that the net effect is large deviation from linearity of the band edges of 
the alloy. Mathematically one diagonalizes an extended Hamiltonian and 
accounts for the shifts in the band edges at the \textit{$\Gamma $} and X point in the Brillouin 
zone. Those shifts are used to calculate the bowing parameters for each 
self-energy.

We calculate the bowing parameters of the direct and indirect bandgap around 0 and 1 limits of 
concentration according to BAC model and follow the spirit of the VCA to interpolate 
linearly the effect of BAC model between these limits. The linear interpolation has been successfully used to 
fit experimental data for ZnSeTe alloy \cite{Walukiewicz:2003,Walukiewicz:2004}. Linear interpolation for the 
direct and indirect bandgaps is consistent with linear interpolation performed on the tight-binding parameters.
The tight-binding parameters for BeTe and BeSe 
are shown in Table~\ref{tab:table1}. The bowing parameters of the on-site energies for Te-rich and Be-rich limit, 
respectively, are shown in parenthesis.

\begin{table}
\caption{\label{tab:table1}Matrix elements in eV of nearest neighbor \textit{sp}$^{3}s^{*}$ model including 
spin-orbit interaction for BeTe and BeSe. The on-site energies of BeTe have been upgraded by 0.41 eV, the band offset 
between BeTe and BeSe. The notation is according to \citeauthor{Vogl:1983} \cite{Vogl:1983}.
The bowing parameters of the on-site energies for BeTe-rich limit and BeSe-rich limit are shown in parenthesis. For 
a general concentration x the linear interpolation is used. }
\begin{ruledtabular}
\begin{tabular}{cccccccc}
& 
BeTe  \par & 
BeSe \par  \\
\hline
E(s,c)& 
5.112+0.41 (-1.85)& 
5.560 (0.55) \\
E(s,a)& 
-15.401+0.41 (0.84)& 
-14.953 (1.0) \\
E(p,c) & 
4.427+0.41 (-0.6)& 
5.026 (-5.7) \\
E(p,a)& 
-0.299+0.41 (0.5)& 
0.300 (5.8) \\
E(s*,c)& 
30.16+0.41 (1.0)& 
21.666 (1.3) \\
E(s*,a) & 
39.203+0.41 (0.5)& 
24.433 (0.65) \\
V(s,s)& 
-3.303& 
-8.195 \\
V(sc,pa)& 
4.423& 
5.633 \\
V(sa,pc)& 
5.511& 
4.89 \\
V(x,x)& 
0.331& 
1.531 \\
V(x,y)& 
6.362& 
6.324 \\
V(s*a,pc)& 
11.503& 
7.462 \\
V(s*c,pa)& 
3.11& 
4.572 \\
$\Delta _{a}$& 
0.97 (-0.4) & 
0.499 (-0.15) \\
$\Delta _{c}$& 
0& 
0 \\
\end{tabular}
\end{ruledtabular}
\label{tab1}
\end{table}

\begin{figure}
\includegraphics{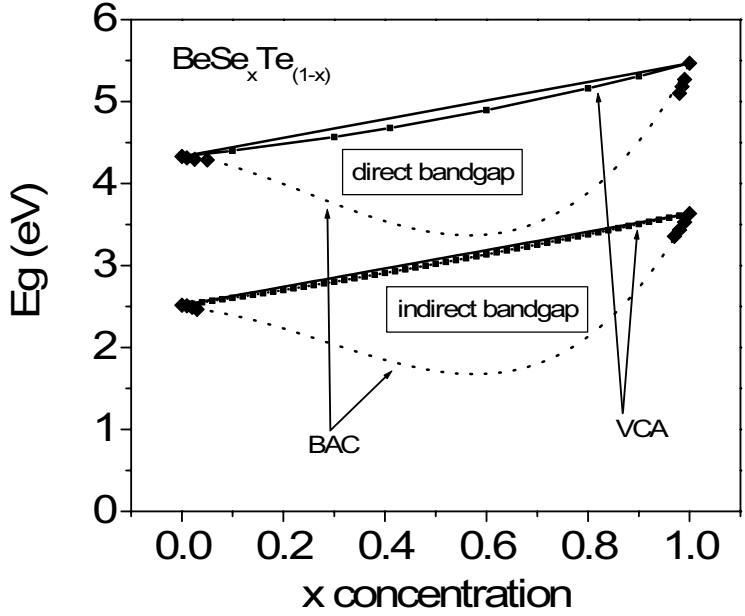}
\caption{\label{fig:1} Direct and indirect bandgaps in BeSe$_{x}$Te$_{1 - x}$ : linear interpolation 
(full line); VCA (line with squares); BAC(diamonds); and interpolation to 
BAC model (dotted line).}
\end{figure}

\noindent
Calculated direct and indirect bandgaps are 
shown in Fig.~\ref{fig:1} for VCA and BAC models against linear interpolation.  The conduction band minimum 
is located at the X-point in the Brillouin zone, such that the fundamental bandgap is indirect. The VCA 
model gives almost constant bowing parameters of 0.49 eV for the direct (optical) bandgap. The bowing of the 
direct bandgap given by the BAC model is much steeper on the Se-rich (9.8 eV) side than on the 
Te-rich side (2 eV) suggesting that utilizing just one bowing parameter is 
inappropriate to describe bandgaps in these structures. The results for the indirect bandgap follow the same 
trend as that of the direct bandgap (Fig.~\ref{fig:1}) with the minimum of the indirect bandgap at 1.7 eV for 
x around 0.6. This trend is similar because of the large bowing of the valence band edge in addition to conduction 
band edge. Moreover, the VCA results are very close to the linear bandgap.

\begin{figure}
\includegraphics{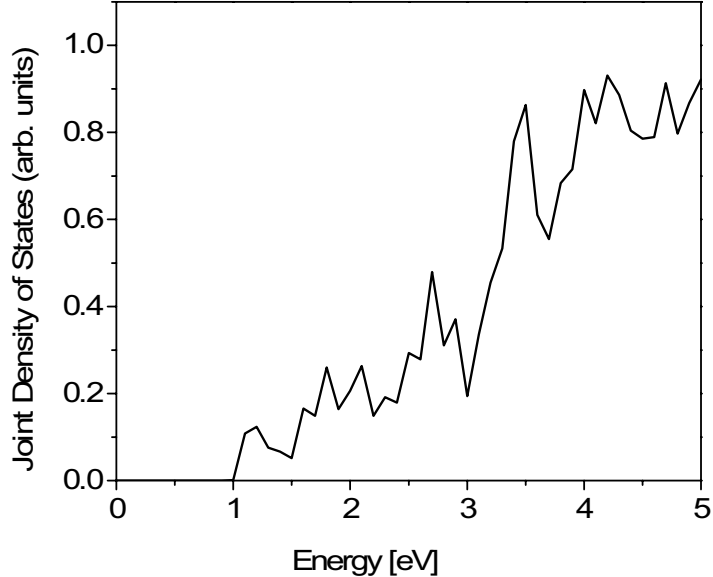}
\caption{\label{fig:2} Joint density of states responsible for vertical 
transitions in Si/BeSe$_{x}$Te$_{1 - x}$ SL.}
\end{figure}

We calculated also electronic and optical properties of a 
Si/BeSe$_{0.41}$Te$_{0.59}$ superlattice (SL) in (001) direction. Abrupt 
interface, flat band conditions were assumed. We adjusted the conduction 
band offset between BeSe$_{0.41}$Te$_{0.59}$  and Si at 1.2 eV as determined from 
electrical measurements \cite{Clark:2000}. Two interface subbands were 
found, one empty and one occupied within the Si bandgap. The origin of these 
interface subbands is due to polar nature of the interface or large 
difference between on-site energies of Si on the one side and Be or Se/Te on 
the other side.\cite{Saito:1992} In Fig.~\ref{fig:2} we plot the joint density of 
states for vertical transitions in 
(Si$_{2})_{4}$(BeSe$_{0.41}$Te$_{0.59})_{4}$ SL. In such structures, due 
to band folding, the threshold of the direct bandgap is lowered for Si from 
3.35 eV to 1 eV. The threshold is slightly below the fundamental bandgap of 
Si because of the two interface bands. Moreover, the first peak is wider. Si 
as an indirect bandgap semiconductor has two kinds of confined states in a 
quantum well \cite{Sandu:2001}, one given by the longitudinal valleys 
with an effective mass of 0.19m$_{0}$ and the other given by transverse 
valley with an effective mass of 0.91m$_{0}$. Hence the confined states that 
have a different first peak is made of the first states of both types of 
valley.

\section{Conclusions}
In conclusion, we extended the band anticrossing model to the empirical 
tight-binding theory. We used the Anderson model for impurity and a sp$^{3}$ 
hybrid basis for zincblende structures within sp$^{3}$s* Hamiltonian. The 
effective Hamiltonian of the alloy was obtained by impurity averaging and 
keeping only the terms responsible for energy shifts due to alloying. The 
pair impurity effects can easily be included as well. Thus no extra 
parameters are needed to calculate bandgaps. 

The model was used for BeSe$_{1 
- x}$Te$_{x}$ alloy. BeSe$_{1 - x}$Te$_{x}$ shows large band bowing, larger 
on the Se-rich side similar to ZnSeTe system. Bandgap was interpolated 
linearly between the two dilute limits. Further the model was applied to 
Si/BeSe$_{0.41}$Te$_{0.59}$ superlattice in (001) direction with abrupt 
interfaces. Due to polarity of the interface, two interface subbands are 
found within bandgap of Si. Also calculations show that the threshold for 
direct transitions is lowered in Si and that the absorption edge is slightly 
below the Si fundamental bandgap.

\begin{acknowledgments}
The material is based in part upon work supported by NASA under awards no. 
NCC-1-02038 and NCC-3-516 and Office of Naval Research. 
\end{acknowledgments}


\end{document}